\documentclass[aps,amsmath,amssymb,showpacs,preprint]{revtex4}
\usepackage{graphicx,bm,dcolumn}

\begin{document}

\title{A reggeized model for $\bm{\eta}$ and $\bm{\eta'}$ photoproduction}

\author{Wen-Tai Chiang}
\author{Shin Nan Yang}
\affiliation{Department of Physics, National Taiwan University,
             Taipei 10617, Taiwan}

\author{L. Tiator}
\author{M. Vanderhaeghen}
\author{D. Drechsel}
\affiliation{Institut f\"ur Kernphysik, Universit\"at Mainz,
             55099 Mainz, Germany}

\date{\today}

\begin{abstract}
A reggeized model for $\eta$ and $\eta'$ photoproduction on the nucleon is
presented. In this model, $t$-channel vector meson exchanges are described
in terms of Regge trajectories to comply with the correct high energy
behavior. We compare this reggeized model with an isobar model
($\eta$-MAID), where the $t$-channel exchanges are described by $\rho$ and
$\omega$ poles. Both models contain the same resonance contributions, and
describe current $\gamma p \rightarrow \eta p$ data up to
$E_\gamma^\mathrm{lab}=$ 2~GeV quite well, but only the reggeized model can
be successfully extended to higher energies. For the $\gamma p \rightarrow
\eta' p$ reaction, the reggeized model is found to be able to give a
satisfactory description. For the differential cross section data from
SAPHIR, we find that the observed linear forward rise in $\cos\theta$ near
$E_\gamma^\mathrm{lab}=$ 1.6~GeV can be well described by the interference
of an $S_{11}$ resonance and the Regge trajectory exchanges without any
need for an additional $P$-wave resonance.
\end{abstract}


\maketitle

\section{Introduction}
\label{sec:Intro}%
Photoproduction of $\eta$ and $\eta'$ on the nucleon, $\gamma N \rightarrow
\eta N,\;\eta' N$, provides an alternative tool to study nucleon resonances
($N^*$) besides $\pi N$ scattering and pion photoproduction. Both the $\eta
N$ and $\eta' N$ states couple to nucleon resonances with isospin $I = 1/2$
only. Therefore, these processes are cleaner and more selective to
distinguish certain resonances than other processes, e.g., pion
photoproduction. This provides the opportunity to access less explored
nucleon resonances, especially some higher mass $N^*$ about which only
little information is as yet available.

During the last decade, the $\eta$ photoproduction has been actively
studied both theoretically and experimentally. The status was reviewed in
our recent study~\cite{Chiang:2001as}. On the other hand, the data on
$\eta'$ photoproduction are still very limited. In 1968, the ABBHHM
collaboration~\cite{unknown:1968ke} observed 11 events in a bubble chamber
using an untagged photon beam. With tagged photons an experiment at DESY in
1976~\cite{Struczinski:1976ik} found approximately 7 $\eta'$ candidates
with a streamer chamber setup. In 1998 the SAPHIR collaboration at
Bonn~\cite{Plotzke:1998ua} explored the energy region from 0.9 to 2.6 GeV
with tagged photons and obtained angular distributions in 7 energy bins.
Unlike $\eta$ photoproduction with an almost constant differential cross
section in the threshold region, the $\eta'$ photoproduction exhibits a
sizeable $P$-wave contribution from threshold to the maximum energy. In an
isobar analysis the data could be well described by two resonances,
$S_{11}$ and $P_{11}$, at energies of $W = 1897$ and $1986$ MeV,
respectively~\cite{Plotzke:1998ua}.

Among the theoretical studies on $\eta'$ photoproduction, Zhang {\it et
al.}~\cite{Zhang:1995uh} used a field theoretical Lagrangian model and
explained the pre-1998 data of total cross sections well. Their
calculations included Born terms, vector meson exchange and resonance
excitation, and strongly emphasized the $D_{13}(2080)$ excitation as the
main production mechanism. Also Li~\cite{Li:1997} could give a good
description of the total cross section data within a constituent quark
model by generating nucleon resonances in the $s$-channel. The dominant
contribution was found from off-shell $S_{11}(1535)$ excitation. In a
similar calculation in 2001, Zhao~\cite{Zhao:2001kk} could also well
describe the SAPHIR data by introducing further $P_{13}$ and $F_{15}$
resonances. In both quark model approaches, $t$-channel vector meson
exchange was not included. In an equally good description of the data,
Borasoy~\cite{Borasoy:2000qv} applied U(3) baryon chiral perturbation
theory with Born terms, vector mesons and off-shell contributions from
$P_{11}(1440)$ and $S_{11}(1535)$ resonances.

A very interesting subject in $\eta$ and $\eta'$ photoproduction is the
meson coupling to the nucleon which determines the Born term contribution.
Already in the SU(3) limit the coupling strength $g_{\eta NN}^2 / 4\pi =
0.8 - 1.9$ is much smaller than for pions ($g_{\pi NN}^2 / 4\pi = 14.3$),
but in an analysis of the angular distributions of eta photoproduction an
even smaller value of $g_{\eta NN}^2 / 4\pi = 0.4 \pm 0.2$ was
determined~\cite{Tiator:1994et}. Such a small value was later explained
within a chiral Lagrangian
approach~\cite{Kirchbach:1996kw,Neumeier:2000fb}, and in a very recent fit
within a chiral constituent quark model a value of only $0.04$ has been
obtained~\cite{Saghai:2001yd}. For the $\eta'$ coupling to the nucleon the
situation is even less clear. Zhang {\it et al.}~\cite{Zhang:1995uh}
applied a quark-model mixing relation with a singlet to octet mixing angle
of $\theta=20^o$ and obtained the relation $g_{\eta' NN} \simeq 0.7\;
g_{\eta NN}$. From forward nucleon-nucleon scattering using dispersion
relations, Grein {\it et al.}~\cite{Grein:1980nw} estimated that both
$g_{\eta NN}^2 / 4\pi$ and $g_{\eta'NN}^2 / 4\pi$ are smaller than $1$, and
in an analysis of the strangeness content of the proton
Hatsuda~\cite{Hatsuda:1990bi} obtained values for the coupling constant in
the range $-3 \le g_{\eta' NN} \le +2$ or $g_{\eta' NN}^2 / 4\pi \le 0.7$.

Previously, we used an isobar model ($\eta$-MAID)~\cite{Chiang:2001as} to
study the $\eta$ photo- and electroproduction on the nucleon, which
described the experimental data quite well. Since both $\eta$ and $\eta'$
have the same quantum numbers, we expected that an extension of the
$\eta$-MAID formalism to the $\eta'$ photoproduction should be
straightforward. However, this was not found to be the case. The reason for
this problem turned out to be the much higher threshold for $\eta'$
compared to $\eta$ production ($W$ = 1896~MeV vs. $W$ = 1486~MeV). The
approach used in the $\eta$-MAID model, which is intended for the resonance
region at about $W\lesssim 2$~GeV, therefore has to be modified as the
energy increases. The main modifications refer to the treatment of the
$t$-channel contributions. The vector meson (e.g., $\rho$ and $\omega$)
exchanges in the $t$-channel are usually included in studies of meson
photoproduction, and calculated by using Feynman (pole-like) propagators.
As the energies increase, $t$-channel form factors are usually needed to
regulate the $t$-channel contributions. However, at very high energies ($W
\gg 2$~GeV) the use of meson poles in the $t$-channel is found to fail.

On the other hand, it is well known that the Regge theory is successful in
describing various reactions at high energy and low momentum transfer. In
Ref.~\cite{Guidal:1997hy}, the Regge trajectories in the $t$-channel have
been applied to pion and kaon photoproduction at high energies with
success. Therefore, in this study we adapt a similar treatment for the
$t$-channel vector meson exchanges and apply it to $\eta$ and $\eta'$
photoproduction.

In Section~2, we describe the model ingredients, and focus on the Regge
trajectory exchanges in the $t$-channel. Our results and a comparison with
both $\eta$ and $\eta'$ photoproduction data are given in Section~3,
followed by a summary and conclusions.

\section{Model}
\label{sec:Model}%
Previously, we have used an isobar model ($\eta$-MAID)~\cite{Chiang:2001as}
to study the $\eta$ photo- and electroproduction on the nucleon. This model
contains Born terms, $t$-channel vector meson exchanges, and nucleon
resonances. Although this model describes the current experimental data in
the resonance region ($W\lesssim 2$~GeV) quite well, it cannot easily be
extended to higher energies because of the vector meson poles used in the
$t$-channel exchanges. To improve this situation, we adopt Regge
trajectories to describe these $t$-channel exchanges. In the next sections,
we first briefly introduce the resonance contributions and Born terms used
in both models (details can be found in Ref.~\cite{Chiang:2001as}); then we
focus on the vector meson exchanges in the $t$-channel.

\subsection{Resonance contributions}%
\label{Regge}%
For the resonance contributions of both the $\eta$-MAID and the reggeized
model, the relevant multipoles $\mathcal{M}_{\ell\pm}$ ($=E_{\ell\pm},\,
M_{\ell\pm}$) are assumed to have a Breit-Wigner energy dependence of the
form
\begin{equation} \label{eq:BWres}
 \mathcal{M}_{\ell\pm}(W) = \tilde{\mathcal{M}}_{\ell\pm}\,
 f_{\gamma N}(W)\,
 \frac{M_R \Gamma_\mathrm{tot}(W)}{M_R^2-W^2-i M_R \Gamma_\mathrm{tot}(W)}\,
 f_{\eta N}(W)\, C_{\eta N} \,,
\end{equation}
where $f_{\eta N}(W)$ is the usual Breit-Wigner factor describing the $\eta
N$ decay of the $N^*$ resonance with total width $\Gamma_\mathrm{tot}$,
partial width $\Gamma_{\eta N}$ and spin $J$,
\begin{equation} \label{eq:fetaN}
 f_{\eta N}(W) = \zeta_{\eta N} \left[ \frac{1}{(2J+1)\pi}\,
 \frac{|\bm{k}|}{|\bm{q}|}\, \frac{M_N}{M_R}\,
 \frac{\Gamma_{\eta N}}{\Gamma_\mathrm{tot}^2} \right]^{1/2},
\end{equation}
with $|\bm{k}|$ and $|\bm{q}|$ the photon and $\eta$ meson momenta in the
c.m. system, and $\zeta_{\eta N} = \pm 1$ a relative sign between the $N^*
\rightarrow \eta N$ and $N^* \rightarrow \pi N$ couplings.

The energy dependence of the partial width $\Gamma_{\eta N}$ is given by
\begin{equation} \label{eq:GametaN}
 \Gamma_{\eta N}(W) = \beta_{\eta N }\,\Gamma_R
 \left(\frac{|\;\bm{q}\;|}{|\bm{q}_R|}\right)^{2\ell+1}
 \left(\frac{X^2+\bm{q}_R^2}{X^2+\bm{q}^2}\right)^\ell \frac{M_R}{W}\,,
\end{equation}
where $X$ is a damping parameter for all resonances, assumed to be
$500$~MeV in the $\eta$-MAID and $450$~MeV in the reggeized model;
$\Gamma_R$ and $q_R$ are the total width and the $\eta$ c.m. momentum at
the resonance peak ($W=W_R$), respectively, and $\beta_{\eta N}$ is the
$\eta N$ decay branching ratio. The total width $\Gamma_\mathrm{tot}$ in
Eqs.~(\ref{eq:BWres}) and (\ref{eq:fetaN}) is the sum of $\Gamma_{\eta N}$,
the single-pion decay width $\Gamma_{\pi N}$, and the rest, for which we
assume dominance of the two-pion decay channels,
\begin{equation}
 \Gamma_\mathrm{tot}(W) =
 \Gamma_{\eta N}(W) + \Gamma_{\pi N}(W) + \Gamma_{\pi\pi N}(W)\,.
\end{equation}
The widths $\Gamma_{\pi N}$ and $\Gamma_{\pi\pi N}$ are parameterized
similarly as for $\Gamma_{\eta N}$. More details about the parameterization
of Eq.~(\ref{eq:BWres}) can be found in Ref.~\cite{Chiang:2001as}.

In our previous study~\cite{Chiang:2001as}, we did not require the form
factor $f_{\gamma N}(W)$ for the $\gamma NN$ vertex in the Breit-Wigner
form Eq.~(\ref{eq:BWres}). However, we now found that it is necessary to
include $f_{\gamma N}(W)$ to describe the data at $W>2$~GeV. Therefore, we
assume the form
\begin{equation} \label{eq:fgammaN}
 f_{\gamma N}(W) =
 \left(\frac{|\;\bm{k}\;|}{|\bm{k}_R|}\right)^2
 \left(\frac{X^2+\bm{k}_R^2}{X^2+\bm{k}^2}\right)^2
\end{equation}
in Eq.~(\ref{eq:BWres}) for all $N^*$. For $S_{11}$(1535) resonance, the
factor $f_{\gamma N}(W)$ decreases its contribution at high energies, but
affects very little near the $S_{11}$(1535) resonance region. For other
resonances, the differences caused by this factor are negligible.

\subsection{Born terms}
\label{sec:Born}%
The nonresonant background contains the usual Born terms and vector meson
exchange contributions. It is obtained by evaluating the Feynman diagrams
derived from an effective Lagrangian. In the $\eta$-MAID model, the Born
terms are constructed in the standard way, and the details can be found in
Ref.~\cite{Chiang:2001as}. In the reggeized model, however, we do not
include the Born terms. The reason is that the correct treatment for the
$u$-channel nucleon exchange, together with the reggeized $t$-channel
vector meson exchanges, requires to also introduce the nucleon Regge
trajectories. Because of the lack of high energy data at backward angles,
it is currently difficult to determine this $u$-channel contribution. Since
the coupling constants $g_{\eta NN}$ and $g_{\eta' NN}$ are small, the
difference caused by the absence of the Born terms is negligible at low
energies. Their effects become important only when the energies increase
(see Fig.~\ref{fig:eta-dxs} and the discussion in Sec.~\ref{sec:result}).

\begin{figure}
\includegraphics[width=0.35\columnwidth]{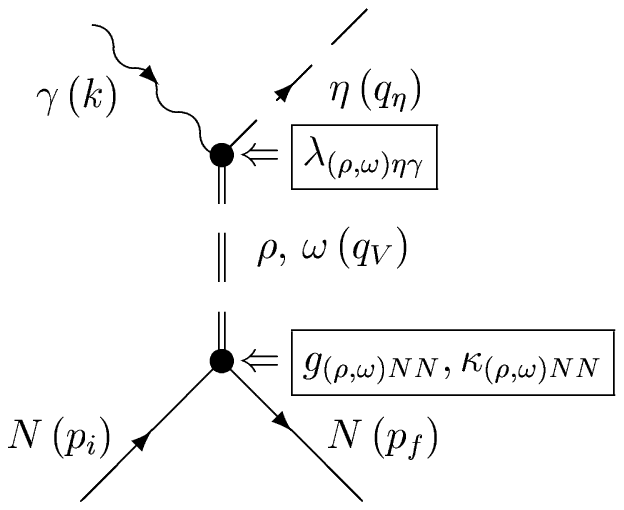}
\caption{\label{fig:vmxdiag}The $t$-channel $\rho$ and $\omega$ meson
exchange diagram in $\eta$ photoproduction. }
\end{figure}

\subsection{Vector meson exchanges in the $\bm{t}$-channel}
\label{sec:tch}%
The Feynman diagram corresponding to $t$-channel vector meson
($V=\rho\,,\;\omega$) exchanges in $\eta$ photoproduction is shown in
Fig.~\ref{fig:vmxdiag}. The electromagnetic coupling constants
$\lambda_{V\eta\gamma}$ and $\lambda_{V\eta'\gamma}$ can be determined from
the radiative decay widths,
\begin{eqnarray}
 \Gamma_{V \rightarrow \eta\gamma}
 &=& \frac{\alpha(m_V^2-m_\eta^2)^3}{24\,m_V^3\,m_\eta^2}\,
     \lambda_{V\eta\gamma}^2 \,,
\\ \label{eq:etapdecay}
 \Gamma_{\eta' \rightarrow V\gamma}
 &=& \frac{\alpha(m_{\eta'}^2-m_V^2)^3}{24\,m_V^2\,m_{\eta'}^3}\,
     \lambda_{V\eta'\gamma}^2 \,,
\end{eqnarray}
where $\alpha$ is the fine-structure constant. In Table~\ref{tbl:vecmes},
we list the values of $\lambda_{V\eta\gamma}$ and $\lambda_{V\eta'\gamma}$
as obtained from the widths of Ref.~\cite{PDG:2000}: $\Gamma_{\rho
\rightarrow \eta\gamma} = 36\ \mathrm{keV}$, $\Gamma_{\omega \rightarrow
\eta\gamma} = 5.5\ \mathrm{keV}$, $\Gamma_{\eta' \rightarrow \rho\gamma} =
89\ \mathrm{keV}$, and $\Gamma_{\eta' \rightarrow \omega\gamma} = 9.1\
\mathrm{keV}$.

\begin{table}
\caption{\label{tbl:vecmes}Parameters for the vector mesons in this study.}
\begin{tabular*}{\columnwidth}{c @{\extracolsep{\fill}} c c c c c c}
\hline
 $V$ & $m_V\,[{\mathrm MeV}]$ & $g_{VNN}$ & $\kappa_{VNN}$
     & $\lambda_{V\eta\gamma}$ & $\lambda_{V\eta'\gamma}$ & $\alpha_V(t)$ \\
\hline
 $\rho$   & $768.5$ & $2.4$ & $3.7$ & $0.81$ & $\ 1.24$
          & $0.55+0.8\,t/\mathrm{GeV}^2$ \\
 $\omega$ & $782.6$ & $9  $ & $0  $ & $0.29$ & $-0.43$
          & $0.44+0.9\,t/\mathrm{GeV}^2$ \\
\hline
\end{tabular*}
\end{table}

For the hadronic $VNN$ vertex, a dipole form factor is included in the
$\eta$-MAID. We choose the same form as in Ref.~\cite{Chiang:2001as}:
$(\Lambda_V^2 - m_V^2)^2 / (\Lambda_V^2 + \bm{q}_V^2 )^2$, with cut-offs
$\Lambda_\rho = 1.0$~GeV and $\Lambda_\rho = 1.3$~GeV from our fit.
However, this hadronic form factor is not required in the reggeized model.
Various values of the hadronic couplings $g_{VNN}$ and $\kappa_{VNN}$ can
be found in the literature. Unlike the $\eta$-MAID where the values of the
$g_{VNN}$ and $\kappa_{VNN}$ couplings are treated as fitting parameters,
the reggeized model contains these hadronic couplings as derived by a fit
to high energy data, which will be discussed later in this section.

The adoption of Regge trajectories for vector meson ($\rho$ and $\omega$)
exchanges allows us to describe the high energy behavior. More details
about applying Regge trajectories for meson photoproduction can be found in
Ref.~\cite{Guidal:1997hy}, which deals with pion and kaon photoproduction
at high energies.

The idea behind the replacement of the pole-like Feynman propagator by a Regge
propagator, is to economically take into account the exchange of high-spin
particles in the $t$- (or $u$-) channels which cannot be neglected any more
as one moves to higher energies.

\begin{figure}
\includegraphics[width=0.50\columnwidth]{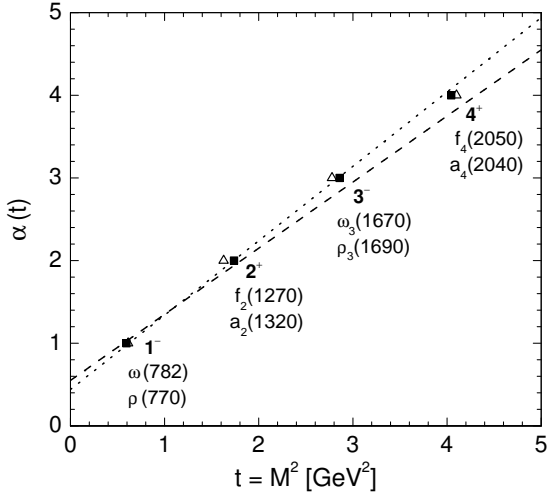}
\caption{\label{fig:Reggetraj}Regge trajectories of the $\rho$ and $\omega$
mesons shown by the dashed and dotted lines, respectively. Mesons
belonging to the $\rho$ ($\omega$) trajectories are indicated by the
symbols $\blacksquare$ ($\triangle$). }
\end{figure}

The Regge trajectories are of the form $\alpha(t) = \alpha_0 + \alpha'\,t$,
where $t$ is the Mandelstam variable, and shown in Fig.~\ref{fig:Reggetraj}
for $\rho$ and $\omega$ trajectories. The numerical values of $\alpha_0$
and $\alpha'$ are taken from Ref.~\cite{Guidal:1997hy}, and given in
Table~\ref{tbl:vecmes}.

$\rho\ (1^-)$ exchange:
\begin{equation} \label{eq:rhoregge}
  \frac{1}{t-m_\rho^2}\ \Longrightarrow\
  \mathcal{P}^\rho_\mathrm{Regge} =
  \left(\frac{s}{s_0}\right)^{\alpha_\rho(t)-1}
  \frac{\pi\alpha'_\rho}{\sin(\pi\alpha_\rho(t))}
  \frac{\mathcal{S}_\rho+e^{-i\pi\alpha_\rho(t)}}{2}
  \frac{1}{\Gamma(\alpha_\rho(t))} \,.
\end{equation}

$\omega\ (1^-)$ exchange:
\begin{equation} \label{eq:omegaregge}
  \frac{1}{t-m_\omega^2}\ \Longrightarrow\
  \mathcal{P}^\omega_\mathrm{Regge} =
  \left(\frac{s}{s_0}\right)^{\alpha_\omega(t)-1}
  \frac{\pi\alpha'_\omega}{\sin(\pi\alpha_\omega(t))}
  \frac{\mathcal{S}_\omega+e^{-i\pi\alpha_\omega(t)}}{2}
  \frac{1}{\Gamma(\alpha_\omega(t))} \,.
\end{equation}

The parameter $s_0$ is a mass scale taken as $s_0 = 1\;\mathrm{GeV}^2$, and
$\mathcal{S} =\pm 1$ is the signature of the trajectory. The gamma function
$\Gamma(\alpha(t))$ suppresses poles of the propagator in the unphysical
region. As is well known from Regge theory~\cite{Collins:1977jy},
trajectories can be either non-degenerate or degenerate. A degenerate
trajectory is obtained by adding or subtracting the two non-degenerate
trajectories (corresponding in our case with $1^-$, $3^-$, $5^-$, ... and
$2^+$, $4^+$, $6^+$, ... states respectively on the $\rho$ or $\omega$
trajectories) with the two opposite signatures. This leads to trajectories
with either a rotating ($e^{-i\pi\alpha(t)}$) or a constant (1) phase. In
line with the finding of Ref.~\cite{Guidal:1997hy} for the $\rho$ and
$\omega$ trajectories, we use the rotating phase in the following. We
further note that the Regge propagator reduces to the Feynman propagator
$1/(t-m^2)$ if one approaches the first pole on a trajectory (i.e., $t
\rightarrow m^2$).

\begin{figure}
\includegraphics[width=0.6\columnwidth]{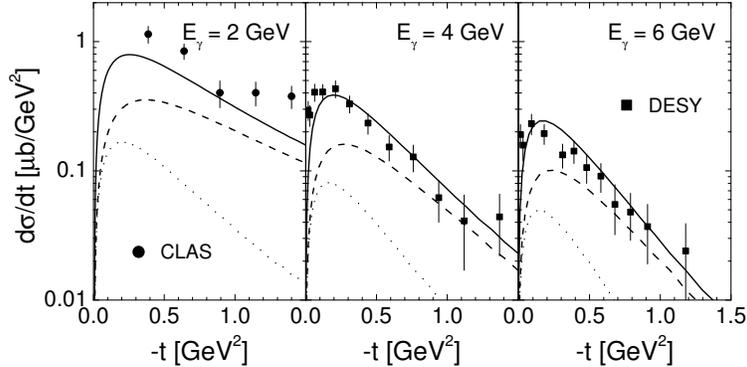}
\caption{\label{fig:etadesy-dsdt}Differential cross section $d\sigma/dt$
for $\gamma p \rightarrow \eta p$. The solid lines are the predictions from
$t$-channel exchange using Regge trajectories, the dashed (dotted) lines
indicate the $\rho$ ($\omega$) contributions only.
The data at $E_\gamma^\mathrm{lab}=$ 4~GeV and 6~GeV are from
DESY~\cite{Braunschweig:1970jb}, at the lower energy we compare with the
CLAS data~\cite{Dugger:2002} at $E_\gamma^\mathrm{lab}=$ 1.925~GeV. }
\end{figure}

Differential cross section data for $\gamma p \rightarrow \eta p$ at
high-$s$ ($E_\gamma^\mathrm{lab}=$ 4 and 6~GeV) and low-$t$ (forward
angles) were measured at DESY~\cite{Braunschweig:1970jb}, as shown in
Fig.~\ref{fig:etadesy-dsdt}. The data can be well described by the
$t$-channel Regge trajectory exchanges . Fitting these data, we determine
the values of the hadronic couplings $g_{VNN}$ and $\kappa_{VNN}$, as given
in Table~\ref{tbl:vecmes}. These values are then fixed and used for our
calculation of both $\eta$ and $\eta'$ photoproduction.

\section{Results and discussion}
\label{sec:result}

\subsection{$\bm{\eta}$ photoproduction results}
\label{sec:eta}%
In this section, we present the $\eta$ photoproduction results from the
reggeized model as well as the $\eta$-MAID model. The differences between
the $\eta$-MAID results presented here and in Ref.~\cite{Chiang:2001as} are
that the former include the recent CLAS photoproduction
data~\cite{Dugger:2002} in the fit. In the reggeized model, we replace the
$t$-channel $\rho$ and $\omega$ exchanges used in $\eta$-MAID by the Regge
trajectories while keeping the $N^*$ contributions. Both models are fitted
to current photoproduction data of cross sections from
TAPS~\cite{Krusche:1995nv}, GRAAL~\cite{Renard:2000iv}, and
CLAS~\cite{Dugger:2002} as well as polarized beam asymmetry from
GRAAL~\cite{Ajaka:1998zi}. Note that we did not include the polarized
target asymmetry measured at ELSA (Bonn)~\cite{Bock:1998rk} in our fit for
the reason discussed in Refs.~\cite{Chiang:2001as,Tiator:1999gr}.

\begin{figure}
\includegraphics[width=0.5\columnwidth]{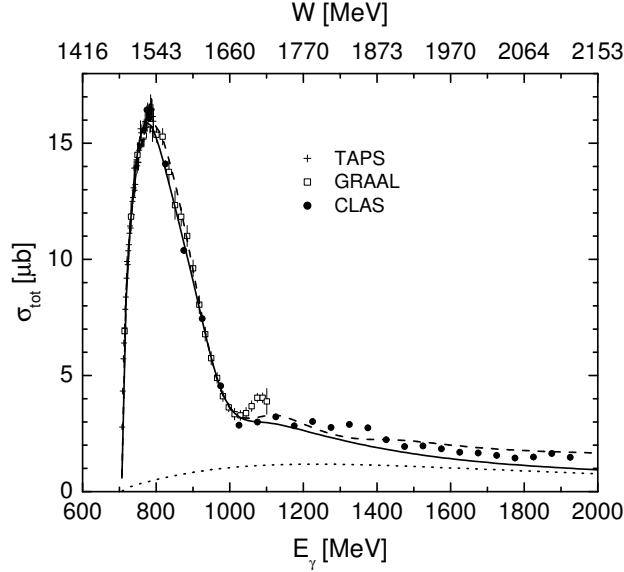}
\caption{\label{fig:eta-txs}Total cross section for $\gamma p \rightarrow
\eta p$. The solid line is the full result from the Regge model, the dotted
line indicates the contribution from $t$-channel Regge exchanges only. The
$\eta$-MAID result is given by the dashed line. Data are from
TAPS~\cite{Krusche:1995nv}, GRAAL~\cite{Renard:2000iv}, and
CLAS~\cite{Dugger:2002}.}
\end{figure}

We show our result for the total cross sections in Fig.~\ref{fig:eta-txs}.
Both models agree with the data quite well. Note that the GRAAL and CLAS
total cross section data shown in Fig.~\ref{fig:eta-txs} are given by
integrating the respective differential cross sections. Since these
measurements have certain limits on angular coverage, extrapolation to
unmeasured regions is inevitable in order to estimate the total cross
sections. Therefore, the obtained total cross sections depend on the
extrapolation procedure. This explains that the bump seen in the GRAAL data
near $W$=1.7~GeV ($E_\gamma^\mathrm{lab} \sim$ 1.1~GeV) does not appear in
the CLAS data. Both data sets agree on the differential cross sections, but
because of the different extrapolations used, they deviate from each other
for the total cross sections.

\begin{figure}
\includegraphics[width=0.6\columnwidth]{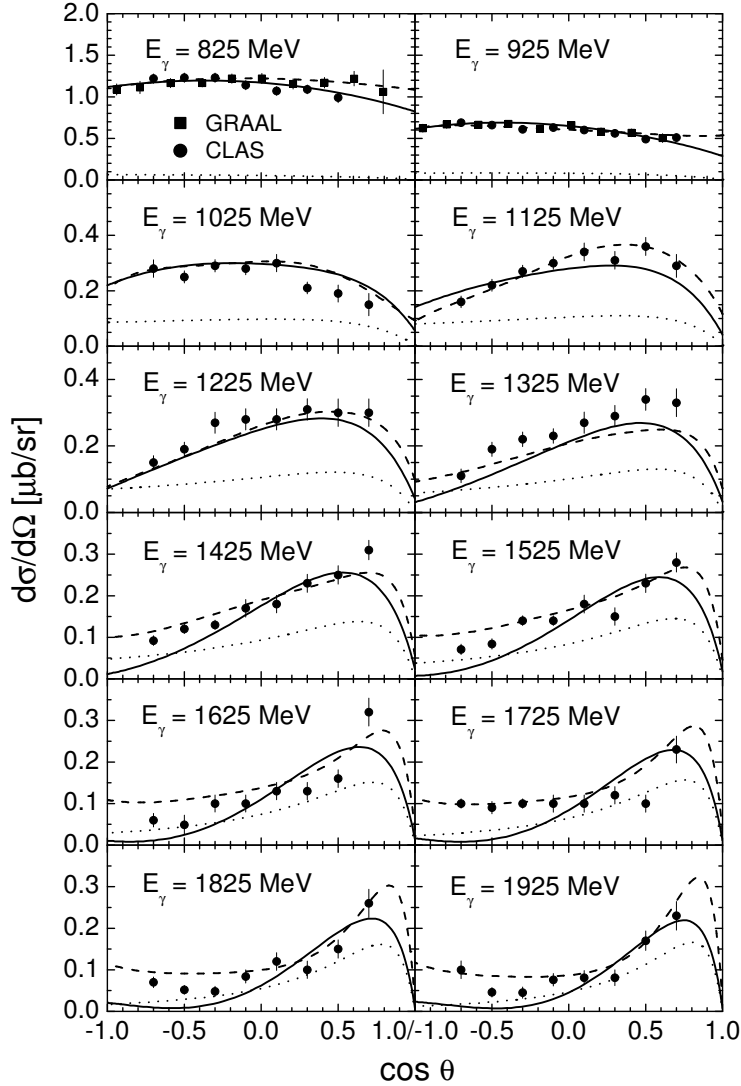}
\caption{\label{fig:eta-dxs}Differential cross section for $\gamma p
\rightarrow \eta p$.
The data are from GRAAL~\cite{Renard:2000iv} and CLAS~\cite{Dugger:2002}.
Notation as in Fig.\ref{fig:eta-txs}.}
\end{figure}

The results for the differential cross sections are given in
Fig.~\ref{fig:eta-dxs}. The overall agreement of the $\eta$-MAID results
with the data is very good. The reggeized model also agrees well with the
data except for an underestimate at backward angles for
$E_\gamma^\mathrm{lab}>$ 1.4~GeV, which probably indicates the influence of
the missing $u$-channel. However, only the reggeized model can be
successfully extended to high energies, as is shown in
Fig.~\ref{fig:etadesy-dsdt} for $E_\gamma^\mathrm{lab}=$ 4 and 6~GeV. We
note that the sharp decrease at forward angles for energies above 1~GeV is
mainly due to the $t$-channel $\rho$ and $\omega$ exchanges.

\begin{figure}
\includegraphics[width=0.6\columnwidth]{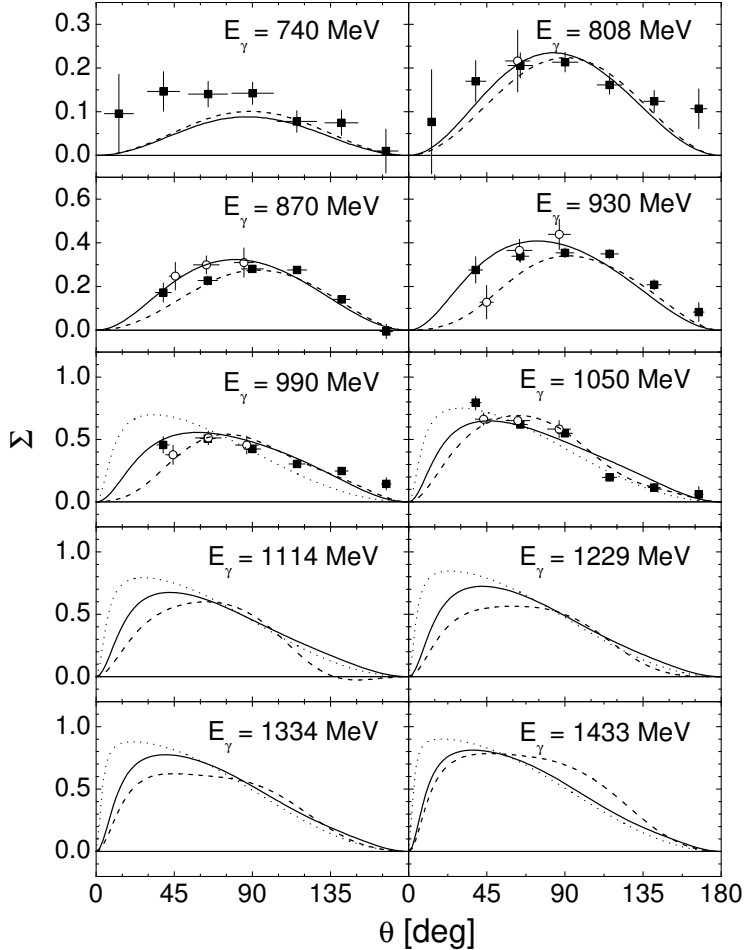}
\caption{\label{fig:eta-beam}Photon beam asymmetry $\Sigma$ for $\gamma p
\rightarrow \eta p$. Notation of the curves as in Fig.~\ref{fig:eta-txs}.
Data are from GRAAL~\cite{Ajaka:1998zi}.}
\end{figure}

The results for the photon beam asymmetry are shown and compared with the
GRAAL data~\cite{Ajaka:1998zi} in Fig.~\ref{fig:eta-beam}. We also include
the preliminary data from GRAAL~\cite{Kuznetsov:nc} for
$E_\gamma^\mathrm{lab}$ between 1.1~GeV and 1.5~GeV in our analysis, but
those data are not shown here. Both models agree with the data reasonably
well. Particularly in the reggeized model, the forward-backward asymmetry
at higher energies is naturally produced by the $t$-channel Regge
trajectory exchanges. The large positive value of the photon asymmetry at
high energies and forward angles indicates the dominance of the Regge
exchanges in this region. Indeed at high $s$ and $-t \ll s$, natural parity
exchange (as in the case of $\rho$ and $\omega$ Regge exchanges) leads to a
photon asymmetry which approaches +1.

\begin{table}
\caption{\label{tbl:respar}Parameters of nucleon resonances studied in the
reggeized model. Masses and widths are given in MeV, $\zeta_{\eta N}$ give
the relative sign between the $N^* \rightarrow \eta N$ and $N^* \rightarrow
\pi N$ couplings, $\beta_{\eta N}$ are the partial decay branching ratios,
and $A^p_{1/2,\,3/2}$ are the photoexcitation helicity amplitudes (in
$10^{-3}\;\mbox{GeV}^{-1/2}$). In the first row for each resonance, we list
the average values or ranges given by the Particle Data Group
(PDG)~\cite{Hagiwara:2002pw}. In the second rows the values are determined
from our fitting; in case of no entry the PDG values are adopted. }
\begin{tabular*}{\columnwidth}{c @{\extracolsep{\fill}} c c c r@{\,\%} c c}
 \hline
 $N^*$  &  Mass  &  Width  &  $\zeta_{\eta N}$
     & \multicolumn{1}{c}{$\beta_{\eta N}$} &  $A^p_{1/2}$ & $A^p_{3/2}$ \\
 \hline
 $D_{13}(1520)$ &          1520  &        120  &      &
                        $0\pm1$  & $- 24\pm 9$ & $+166\pm 5$     \\
                &                &             & $+1$ &
                         $0.04$  &             &                 \\ [1ex]
 $S_{11}(1535)$ &     1520-1555  &    100-250  &      &
                        $30-55$  & $+ 90\pm30$ & ---             \\
                &          1521  &        118  & $+1$ &
                           $50$  &      $+ 80$ & ---             \\ [1ex]
 $S_{11}(1650)$ &     1640-1680  &    145-190  &      &
                         $6\pm1$  & $+ 53\pm16$ & ---             \\
                &          1635  &        120  & $-1$ &
                          $ 16$  &       $+46$ & ---             \\ [1ex]
 $D_{15}(1675)$ &     1670-1685  &        150  &      &
                        $0\pm1$  & $+ 19\pm 8$ & $+ 15\pm 9$     \\
                &          1665  &             & $+1$ &
                         $ 0.7$  &             &                 \\ [1ex]
 $F_{15}(1680)$ &     1675-1690  &        130  &      &
                        $0\pm1$  & $- 15\pm 6$ & $+133\pm12$     \\
                &          1670  &             & $+1$ &
                       $ 0.003$  &             &                 \\ [1ex]
 $D_{13}(1700)$ &          1700  &        100  &      &
                        $0\pm1$  & $- 18\pm13$ & $-  2\pm24$     \\
                &                &             & $-1$ &
                        $ 0.03$  &             &                 \\ [1ex]
 $P_{11}(1710)$ &     1680-1740  &        100  &      &
                        $6\pm1$  & $+  9\pm22$ & ---             \\
                &          1701  &             & $-1$ &
                          $ 26$  &             &                 \\ [1ex]
 $P_{13}(1720)$ &          1720  &        150  &      &
                        $4\pm1$  & $+ 18\pm30$ & $- 19\pm20$     \\
                &                &             & $-1$ &
                           $ 4$  &             &                 \\
 \hline
\end{tabular*}
\end{table}

In Table~\ref{tbl:respar} we present the nucleon parameters extracted from
our fit using the reggeized model. Among the nucleon resonances included in
this study, we found that only the contributions from the $S_{11}(1535)$,
$S_{11}(1650)$, and $D_{13}(1520)$ can be identified unambiguously by this
reaction. The contributions from the other resonances are entangled near
$W=1700$~MeV, and we find it difficult if not impossible to distinguish
their individual contributions from the current $\eta$ photoproduction
data. Comparing with our previous $\eta$-MAID results~\cite{Chiang:2001as},
we find several resonances (e.g., $D_{15}(1675)$ and $F_{15}(1680)$) with
much smaller $\beta_{\eta N}$ in the reggeized model, implying that these
resonance contributions become spurious and redundant once the $t$-channel
exchanges have been reggeized. Furthermore, the respective parameters
should be cautiously interpreted, because several of them (e.g.,
$\Gamma_R$, $\beta_{\eta N}$,\ {\rm{and}}\ $A_{1/2,\,3/2}$) are highly
correlated. Therefore, these studies of eta physics have to be combined
with investigations of other channels or multichannel analyses to give
reliable and convincing information about these resonances.

In conclusion, we find that both the $\eta$-MAID and the reggeized model
give an overall good description of the current data up to
$E_\gamma^\mathrm{lab}$ = 2~GeV. ($\chi^2/N_{\mathrm{dof}} = 2.0$ for the
$\eta$-MAID and $3.9$ for the reggeized fit.) The reggeized model shows
some discrepancies at backward angles because the $u$-channel is not
included, but it is able to describe data at energies as high as
$E_\gamma^\mathrm{lab}=$ 6~GeV, which are beyond the validity of the
$\eta$-MAID model. It is not completely clear at which energy the Regge
trajectories are necessary to describe the $t$-channel. In our present
study of $\eta$ photoproduction, we find that reggeization is not required
for energies up to $W \sim 2$~GeV. However, the situation is different for
$\eta'$ photoproduction. For this reaction, we now show that the Regge
trajectories are required even at energies already near the $\eta'$
production threshold.

\subsection{$\bm{\eta'}$ photoproduction results}
\label{sec:eta'}%
The experimental data basis for $\eta'$ photoproduction is still rather
limited. Besides the total cross section data from
AHHBBM~\cite{unknown:1968ke} and AHHM~\cite{Struczinski:1976ik} measured
decades ago, the only modern data were obtained at SAPHIR
(Bonn)~\cite{Plotzke:1998ua}. However, this status will be largely improved
by new data from CLAS, GRAAL, and CB-ELSA expected to come soon.

The vector meson couplings used in the $t$-channel exchanges are well
determined: the photon couplings $\lambda_{(\rho,\omega)\eta'\gamma}$ can
be obtained from the electromagnetic decay widths of $\eta' \rightarrow
\rho \gamma$ and $\eta' \rightarrow \omega \gamma$ in
Eq.~(\ref{eq:etapdecay}), and for the strong couplings
$g_{(\rho,\omega)NN}$ and $\kappa_{(\rho,\omega)NN}$ the same values as in
$\eta$ photoproduction are used. Furthermore, we find that the Born terms
yield only small contributions because of the small $g_{\eta'NN}$ coupling
suggested by various
studies~\cite{Zhang:1995uh,Grein:1980nw,Hatsuda:1990bi}. The current data
are not able to determine such a small contribution, and thus we do not
include the Born terms as in the case of $\eta$ photoproduction. Therefore,
the background contributions are completely fixed, and only resonance
parameters are varied to fit the data.

\begin{figure}
\includegraphics[width=0.5\columnwidth]{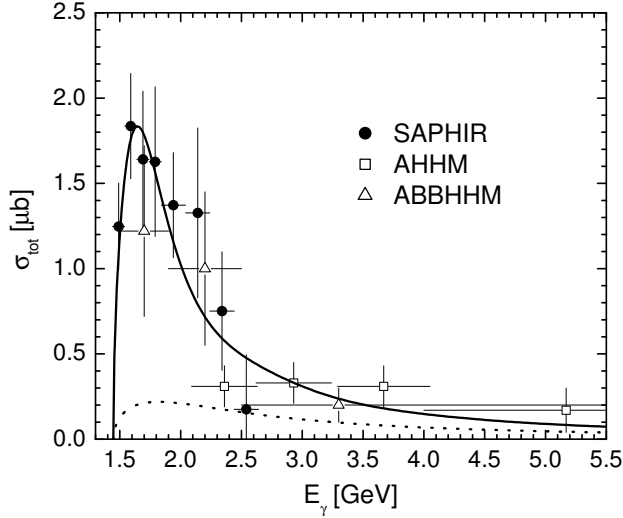}
\caption{\label{fig:etap-regge-txs}Total cross sections for $\gamma p
\rightarrow \eta' p$. The results are obtained by using Regge trajectories
in the $t$-channel exchanges. The solid line shows the full results, and
the dotted line indicates the $t$-channel contribution only. Data are from
SAPHIR~\cite{Plotzke:1998ua}, AHHM~\cite{Struczinski:1976ik}, and
ABBHHM~\cite{unknown:1968ke}.}
\end{figure}

In the total cross section of $\gamma p \rightarrow \eta' p$
(Fig.~\ref{fig:etap-regge-txs}), we observe a sharp rise at threshold and a
quick fall-off with energy. This behavior, also seen in $\eta$
photoproduction, should be due to a strong $S$-wave contribution, most
likely a dominant $S_{11}$ nucleon resonance. Therefore, we start with only
one $S_{11}$ nucleon resonance and $t$-channel Regge trajectory exchanges
to describe the $\gamma p \rightarrow \eta' p$ reaction.

The SAPHIR~\cite{Plotzke:1998ua} data contain the total and differential
cross sections from near the threshold ($E_\gamma^\mathrm{lab}$ = 1.45~GeV)
up to $E_\gamma^\mathrm{lab}$ = 2.44~GeV. The results of our reggeized
model for the total cross section are shown in
Fig.~\ref{fig:etap-regge-txs}, and found to agree well with the SAPHIR
data, along with earlier data from AHHM~\cite{Struczinski:1976ik} and
ABBHHM~\cite{unknown:1968ke}.

\begin{figure}
\includegraphics[width=0.5\columnwidth]{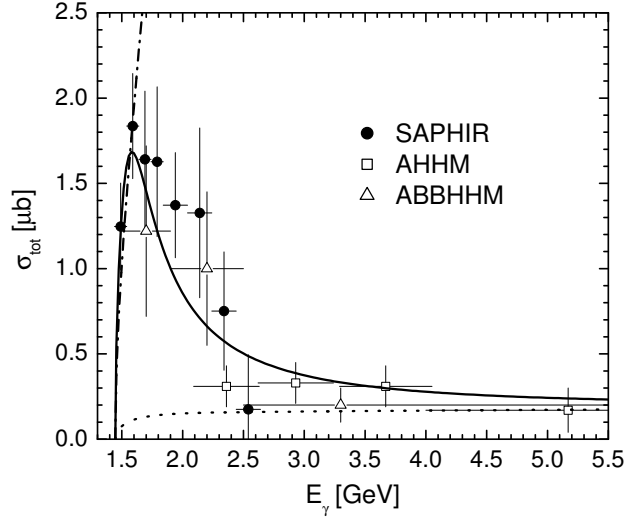}
\caption{\label{fig:etap-pole-txs}Total cross sections for $\gamma p
\rightarrow \eta' p$. These results are obtained by using $\rho$ and
$\omega$ poles in the $t$-channel exchanges, including hadronic form
factors in the $t$-channel. The solid line shows the full results, and the
dotted line indicates the $t$-channel contribution only. The dash-dotted
line indicates the full results without hadronic form factors in the
$t$-channel. Data as in Fig.~\ref{fig:etap-regge-txs}.}
\end{figure}

We also attempt to describe the $\eta'$ photoproduction by using the $\rho$
and $\omega$ poles in the $t$-channel exchanges just as we did in the
$\eta$-MAID model. It is shown in Fig.~\ref{fig:etap-pole-txs} that the
contributions from $\rho$ and $\omega$ pole exchanges without any hadronic
form factors increase drastically above threshold. This can be partly
improved by including hadronic form factors as shown in
Fig.~\ref{fig:etap-pole-txs}. The functional form of these hadronic form
factors is taken the same as in the $\eta$-MAID, with cut-offs
$\Lambda_\rho = 1.1$~GeV and $\Lambda_\rho = 1.5$~GeV from our fit.
Although it is possible to obtain acceptable result for the total cross
sections, this pole description fails to reproduce differential cross
section data, which is discussed next.

\begin{figure}
\includegraphics[width=0.6\columnwidth]{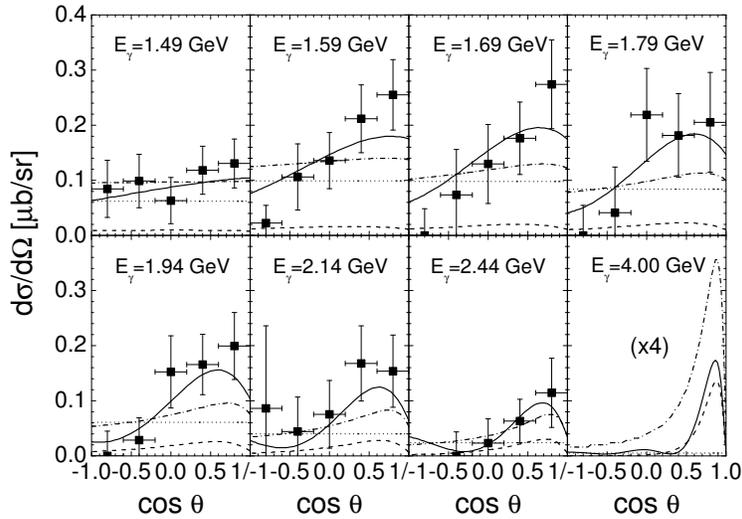}
\caption{\label{fig:etap-dxs}Differential cross section for $\gamma p
\rightarrow \eta' p$. The full results are given by the solid lines, and
the dashed (dotted) lines indicate the contributions from the reggeized
$t$-channel exchanges ($S_{11}$ resonance) only. The dash-dotted lines are
the full results when using $\rho$ and $\omega$ poles instead of Regge
trajectories in the $t$-channel exchanges. Data are from
SAPHIR~\cite{Plotzke:1998ua}.}
\end{figure}

In Fig.~\ref{fig:etap-dxs}, we compare the results for the differential
cross sections with the SAPHIR data~\cite{Plotzke:1998ua} using Regge
trajectories and $\rho$ and $\omega$ poles in the $t$-channel exchanges. We
also give a prediction at $E_\gamma^\mathrm{lab}$ = 4~GeV showing a
pronounced forward peak, which is typical for $\rho$ and $\omega$ Regge
trajectories (and poles) at the higher energies. We observe that the SAPHIR
data show a linear forward rise in $\cos\theta$ at $E_\gamma^\mathrm{lab}$
= 1.59 and 1.69~GeV. Surprisingly, this $P$-wave behavior can be almost
reproduced by our reggeized model, which includes only one $S_{11}$
resonance and $t$-channel exchanges, without introducing a $P$-wave
resonance. The individual contributions from the $S_{11}$ resonance and
$t$-channel exchanges are also shown in Fig.~\ref{fig:etap-dxs}, and both
contributions have a nearly uniform angular distribution in the
differential cross sections at these energies. Therefore, this linear
behavior in $\cos\theta$ is caused by a strong interference between the
$S_{11}$ resonance and $t$-channel Regge trajectory exchanges. This can be
more easily understood if we expand the differential cross sections into a
power series of $\cos\theta$,
\begin{equation} \label{eq:dcs-ps}
 \frac{d\sigma}{d\Omega} = \sum_{\ell=0}^\infty d\sigma_\ell =
 \frac{|\bm{q}|}{|\bm{k}|}
 \sum_{\ell=0}^\infty a_\ell\,\cos^\ell\theta \,.
\end{equation}
where $d\sigma_\ell$ corresponds to the $\cos^\ell\theta$ term, and the
angle-independent coefficients $a_\ell$ can be expressed in terms of
multipoles. If we keep the $S$- and $P$-wave multipoles only and neglect
higher partial waves, then $a_1$ is given as
\begin{equation} \label{eq:a1}
  a_1 = 2\,\mathrm{Re} \left[ E_{0+}^*(3\,E_{1+}+M_{1+}-M_{1-}) \right] \,.
\end{equation}
Therefore, the linear behavior in $\cos\theta$ for differential cross
sections is basically the interference between the $S$- and $P$-wave
multipoles.

\begin{figure}
\includegraphics[width=0.55\columnwidth]{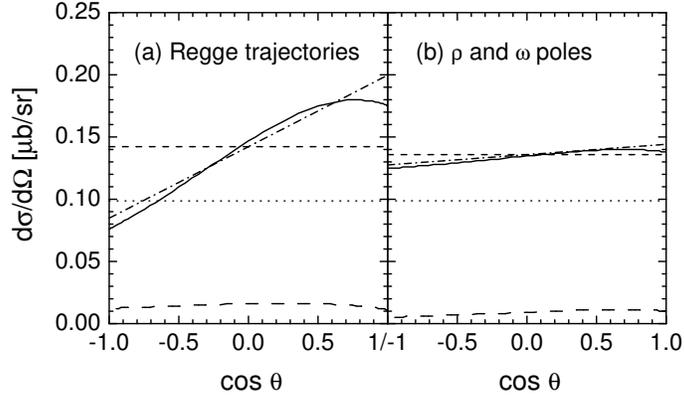}
\caption{\label{fig:dcs-spwave}The results for the differential cross
section $\gamma p \rightarrow \eta' p$ at $E_\gamma^\mathrm{lab}=1.59$~GeV
using (a) Regge trajectories and (b) $\rho$ and $\omega$ poles in the
$t$-channel exchanges. The full results are given by the solid lines, and
the dotted (long dashed) lines indicate the contributions from the $S_{11}$
resonance ($t$-channel exchanges) only. The dashed lines indicate the
uniform angular contribution ($d\sigma_0$) only, while the dash-dotted
lines also include the $S$-$P$ wave interference term
($d\sigma_0+d\sigma_1$).}
\end{figure}

\begin{figure}
\includegraphics[width=0.4\columnwidth]{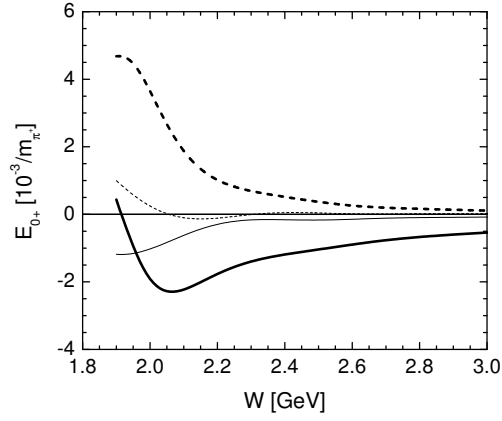}
\caption{\label{fig:regge-E0}$E_{0+}$ multipoles (in $10^{-3}/m_{\pi^+}$)
for $\gamma p \rightarrow \eta' p$. The thick lines are the full result,
the thin lines correspond to Regge exchanges only. The real (imaginary)
parts are indicated by solid (dashed) lines.}
\end{figure}

However, this behavior is not reproduced by the model using the $\rho$ and
$\omega$ poles in the $t$-channel, as this pole description gives almost
flat differential cross section and fails to describe the data (see
Fig.~\ref{fig:etap-dxs}). The reason can be explained in
Fig.~\ref{fig:dcs-spwave}, where we plot the $d\sigma_0$ and $d\sigma_1$
terms (defined in Eq.~(\ref{eq:dcs-ps})) at
$E_\gamma^\mathrm{lab}=1.59$~GeV using (a) Regge trajectories and (b)
$\rho$ and $\omega$ poles in the $t$-channel exchanges for comparison. It
is clearly shown that these two models generate very different $d\sigma_1$
terms, resulting in very distinct differential cross sections. The pole
description generates much smaller $S$- and $P$-wave interference because
the amplitudes from the $\rho$ and $\omega$ poles are real, and contain
fewer higher partial waves. On the contrary, Regge trajectories generate
complex amplitudes with more important contributions from higher partial
waves. As the $S_{11}$ resonance yields dominantly an imaginary
contribution to the $E_{0+}$ multipole around threshold, as can be seen
from Fig.~\ref{fig:regge-E0}, the $S$-$P$ wave interference of
Eq.~(\ref{eq:a1}) is enhanced due to the imaginary part of the $P$-wave
Regge multipoles.

Therefore, our study shows that using Regge trajectories provides much
better description for the $\eta'$ photoproduction than using the $\rho$
and $\omega$ poles. On the other hand, for the $\eta$ photoproduction in
the resonance region a pole description as used in the $\eta$-MAID model
gives a satisfactory description. The pole description may be improved by
introducing additional form factors depending on $s$. However, the
inclusion of an $s$-dependent form factor in a $t$-channel exchange term
would violate gauge invariance, and even so it would still eventually fail
at the higher energies.

\begin{figure}
\includegraphics[width=0.6\columnwidth]{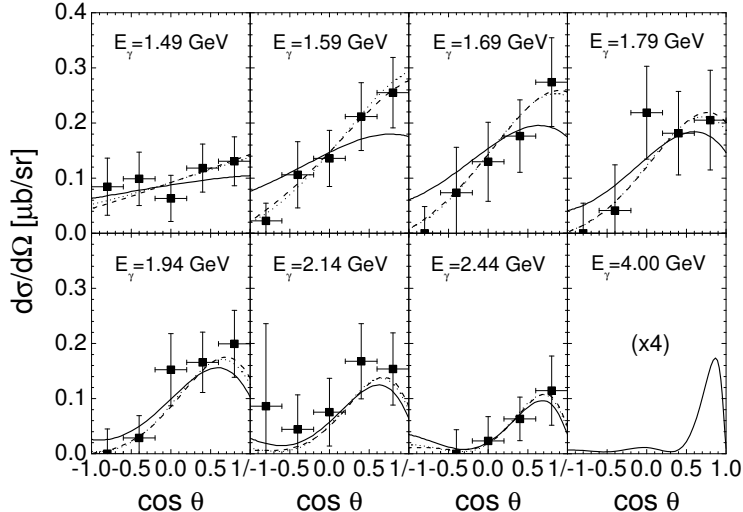}
\caption{\label{fig:etapres-dxs}Differential cross section of $\gamma p
\rightarrow \eta' p$ for various $N^*$ fits. The $S_{11}$ plus reggeized
$t$-channel result is given by the solid line, and the result with an
additional $P_{11}$ ($P_{13}$) is indicated by the dashed (dotted) line.
Data are from SAPHIR~\cite{Plotzke:1998ua}.}
\end{figure}

\begin{figure}
\includegraphics[width=0.55\columnwidth]{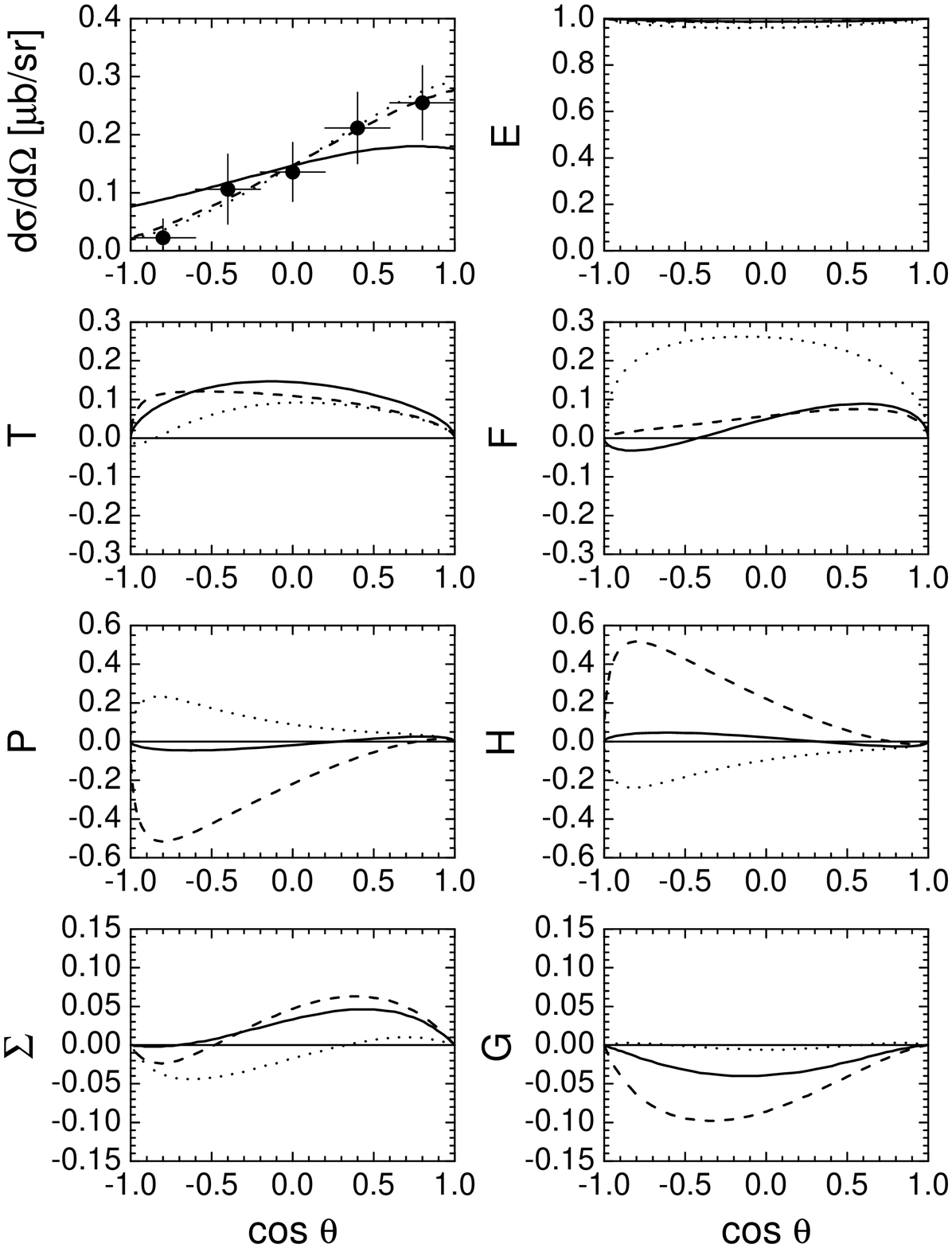}
\caption{\label{fig:etapres-pol}Spin observables for $\gamma p \rightarrow
\eta' p$ at $E_\gamma^\mathrm{lab}=1.59$~GeV for various $N^*$ fits. The
data are from SAPHIR~\cite{Plotzke:1998ua}. Notation as in
Fig.~\ref{fig:etapres-dxs}. }
\end{figure}

Although our model including only the $S_{11}$ resonance and $t$-channel
exchanges fits the data in terms of the overall $\chi^2$
($\chi^2/N_{\mathrm{dof}}=0.64$), the fit to the SAPHIR differential cross
section data at $E_\gamma^\mathrm{lab}$ = 1.59 and 1.69~GeV can be improved
by including an additional $P$-wave nucleon resonance. We find that both a
$P_{11}$ and a $P_{13}$ resonance equally well improve the fit, as shown in
Fig.~\ref{fig:etapres-dxs}. However, even a very precise cross section
measurement cannot distinguish between a $P_{11}$ or $P_{13}$ resonance
contribution without polarization measurements. In
Fig.~\ref{fig:etapres-pol}, we therefore give the predictions for the
single spin observables ($T$, $P$, $\Sigma$) and the beam-target double
spin observables ($E$, $F$, $G$, $H$). We find that the recoil polarization
($P$) is the most sensitive single spin observable to $P_{11}$ or $P_{13}$
admixtures, and that the double spin observables $H$ shows equal
sensitivity.

In Table~\ref{tbl:etapresfit}, we list the $N^*$ masses extracted from the
single $S_{11}$ fit, and the fits with an additional $P_{11}$ or $P_{13}$
resonance. Although none of these resonances is well established, they are
predicted by the constituent quark model and have been found in various
analyses at $W = 1.9 - 2.1$~GeV. In fact the PDG lists $S_{11}(2090)$,
$P_{11}(2100)$, and $P_{13}(1900)$ as one- or two-star
resonances~\cite{Hagiwara:2002pw}.

\begin{table}
\caption{\label{tbl:etapresfit}Results of $N^*$ masses extracted from
various fits for $\gamma p \rightarrow \eta' p$.}
\begin{tabular*}{\columnwidth}{l @{\extracolsep{\fill}} c c c}
\hline
 Resonance fit & $S_{11}$ Mass & $P_{11}$ Mass & $P_{13}$ Mass \\
 \hline
 $S_{11}+t$-channel        & $(1959\pm35)$ MeV &    ---    &    ---      \\
 $S_{11}+P_{11}+t$-channel & $(1932\pm16)$ MeV & (1951$\pm$32) MeV & --- \\
 $S_{11}+P_{13}+t$-channel & $(1933\pm14)$ MeV & --- & $(1954\pm37)$ MeV \\
 \hline
\end{tabular*}
\end{table}

In our model discussed so far, we include the reggeized $t$-channel
exchanges as well as $s$-channel resonances. Phenomenologically, hadronic
scattering amplitudes were found \cite{Dolen68} to exhibit the property of
duality, meaning that the amplitude can be obtained as a sum over
$s$-channel resonances or as a sum over ($t$-channel) Regge pole exchanges.
In particular, it has been shown quantitatively (e.g for forward $\pi^- p
\to \pi^0 n$ scattering) that these amplitudes satisfy finite energy sum
rules, so that in the integral sense the sum over all $s$-channel
resonances yields the same result as the sum over $t$-channel Regge poles,
which is known as global duality. An addition of $s$-channel resonances and
$t$-channel (Regge) exchanges therefore leads to some double counting. A
pioneering model to implement this property of resonance-reggeon duality
for hadronic scattering amplitudes was proposed by
Veneziano~\cite{Veneziano68} and lead to many subsequent works. Such dual
resonance models~\cite{Frampton86} were studied in quite some details and
applied in a variety of processes, in particular to meson-meson scattering
amplitudes. For meson-baryon scattering or meson photoproduction, models
based on the Veneziano model are usually too restrictive to give a
realistic description of the scattering in the resonance region (as they
imply strict relations between the coupling constants appearing in
$s$-channel and $t$-channel processes). For an interesting dual amplitude
model applied to $\pi N$ scattering, which contains Regge asymptotic
behavior for forward and backward angles and which shows resonance features
at low energy, we refer to Ref.~\cite{Alcock75}.

\begin{figure}
\includegraphics[width=0.6\columnwidth]{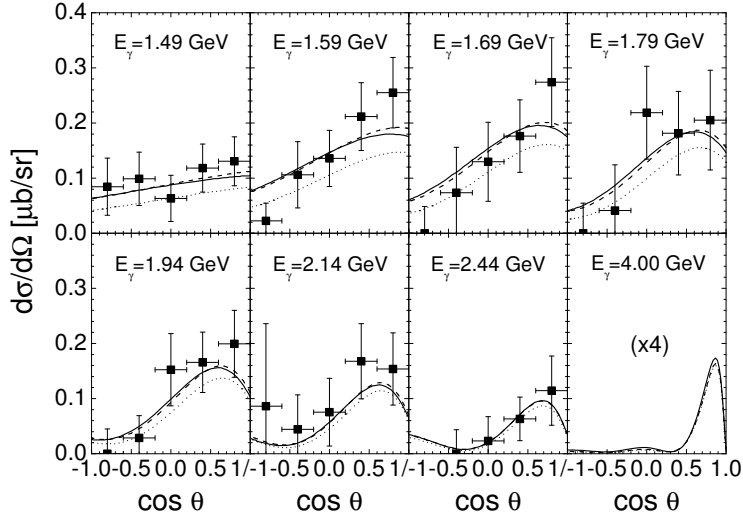}
\caption{\label{fig:etap-noswave-dxs}Differential cross sections for
$\gamma p \rightarrow \eta' p$. The solid line is the same as in
Fig.~\ref{fig:etap-dxs}. The dotted (dashed) line indicates the results
obtained by removing the $S$-wave from the $t$-channel Regge exchanges
before (after) refitting the data. Data as in Fig.~\ref{fig:etap-dxs}.}
\end{figure}

Without claiming that we solve the double counting problem and implement
the property of duality, we study here as a first step a prescription to
demonstrate the order of magnitude of double counting. In
Fig.~\ref{fig:etap-noswave-dxs}, we show the results when we completely
remove the $S$-wave from the $t$-channel Regge trajectory exchanges at all
energies. It appears that after refitting to the data, we essentially
obtain the same results. Of course, the extracted resonance parameters are
affected by this procedure.

Since several parameters (e.g., $A^p_{1/2}$, $\beta_{\eta N}$, and
$\Gamma_\mathrm{tot}$) associated with these resonances are highly
correlated among each other and cannot be well determined individually from
current information, it is not appropriate to discuss the resonance
parameters directly. Instead, we use the quantity
\begin{equation}\label{eq:xi}
  \xi = \sqrt{\frac{m_N k_R}{M_R q_R}
  \frac{\beta_{\eta' N}}{\Gamma_\mathrm{tot}}}\, A^p_{1/2} \,,
\end{equation}
which is not sensitive to uncertainties for individual parameters and thus
less model dependent. Using the parameter $\xi$, we find that its values
before and after the refitting due to removing the $S$-wave are
0.069~$\mathrm{GeV}^{-1}$ and 0.082~$\mathrm{GeV}^{-1}$. Therefore, the
increase of about 20\% in $\xi$ indicates the maximum degree of the double
counting.

\begin{figure}
\includegraphics[width=0.6\columnwidth]{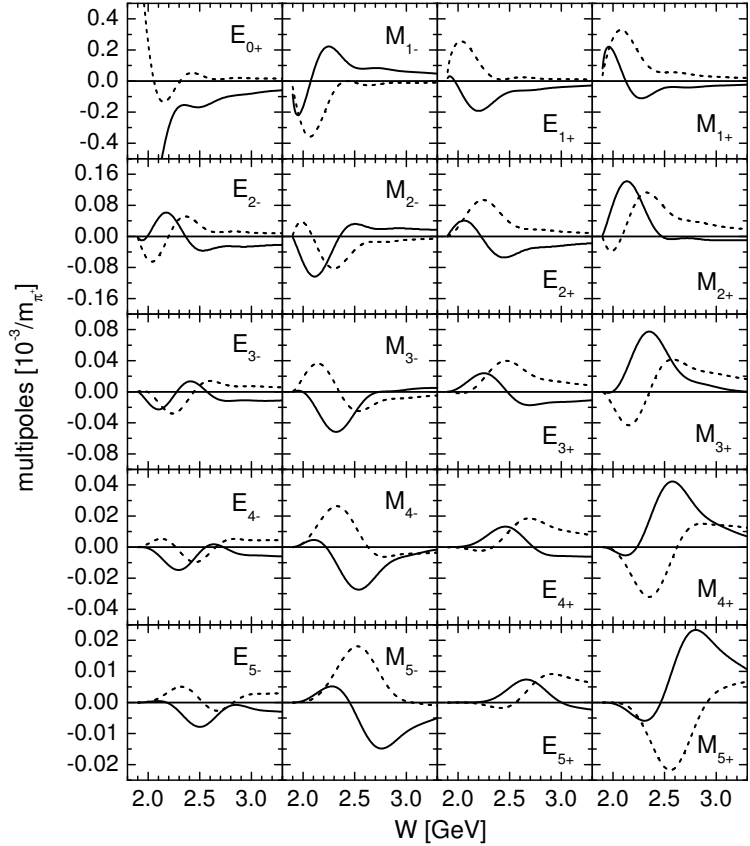}
\caption{\label{fig:regge-mult}Multipoles (in $10^{-3}/m_{\pi^+}$) for the
$t$-channel Regge exchanges in the $\gamma p \rightarrow \eta' p$ process.
Real (imaginary) parts are indicated by solid (dashed) lines.}
\end{figure}

The $E_{0+}$ multipoles predicted from our model are given in
Fig.~\ref{fig:regge-E0}. We note that Regge exchanges yield a finite
imaginary part of $E_{0+}$, while the pole exchanges in the $t$-channel
always give real contributions. In Fig.~\ref{fig:regge-mult} we also show
the multipoles for the $t$-channel Regge exchanges for $l \leq 5$.

\section{Summary and conclusions}
\label{sec:sum}%
In this paper, we present a new study for $\eta$ and $\eta'$
photoproduction using a reggeized model. This model contains resonance
contributions along with reggeized vector meson exchanges in the
$t$-channel, and yields a good description for these reactions in and
beyond the resonance region.

We apply the reggeized model to $\eta$ photoproduction and compare our
results with the $\eta$-MAID model, where $\rho$ and $\omega$ poles are
used instead. We find that both models allow us to achieve reasonably good
fits of the current data from Mainz, GRAAL, and CLAS. Therefore, it is
probably still appropriate to describe the $t$-channel in terms of the
$\rho$ and $\omega$ poles in the resonance regions ($W\lesssim 2$~GeV).
However, only the reggeized model can be successfully applied to describe
the high energy data of the DESY experiment at $E_\gamma^\mathrm{lab}=$ 4
and 6~GeV.

In the case of $\eta'$ photoproduction, we find that its production
threshold is too high to allow an acceptable description with vector meson
poles in the $t$-channel. Unlike the case of $\eta$ photoproduction, the
use of Regge trajectories is able to improve the description. Including
only an $S_{11}$ resonance and the reggeized $t$-channel, we are able to
describe the current data. We find that most of the $P$-wave contributions
are generated by the interference of a $S_{11}$ resonance and $\rho$ and
$\omega$ reggeized exchanges. In this study, we extract an $S_{11}$
resonance with a mass in the range of 1932-1959 MeV, the exact value
depending on whether or not an additional $P$-wave $N^*$ is introduced.
Though this procedure gives some evidence for such an $S_{11}$ resonance,
it does certainly not establish a resonance by itself. The
PDG~\cite{Hagiwara:2002pw} lists the $S_{11}(2090)$ as a one-star
resonance, and quotes previous results from
Refs.~\cite{Manley:1992yb,Cutkosky:1979fy,Hohler:1979yr} where the mass
varies from 1880 to 2180~MeV. Recent analysis of pion-nucleon scattering
and pion photoproduction~\cite{Chen:2002mn} also indicates the existence of
such a resonance. Furthermore, various quark models (e.g.,
Ref.~\cite{Giannini:2002vp}) have predicted an $S_{11}$ resonance in this
energy region. Therefore, the $\eta'$ photoproduction provides a good
channel to study this less explored resonance and possibly other
higher-mass resonances as well.

Occasionally, a tendency of backward peaked $\eta$ and $\eta'$ differential
cross sections was reported.  Though this behavior has not been fully
confirmed, it is likely due to the $u$-channel nucleon exchange and should
be studied in future work by also introducing the nucleon Regge
trajectories in the $u$-channel.

The new forthcoming data for $\eta$ and $\eta'$ photo- and
electroproduction will yield novel and interesting information on the
nucleon resonance region. However, the analysis of these data in terms of
nucleon resonances in the $s$-channel will require a consistent description
of the observables over a large energy range and an improved treatment of
the $t$- and $u$-channel backgrounds by means of reggeized trajectories.
The present work can serve as a first step in this direction.

\begin{acknowledgments}
The authors would like to thank the members of the GRAAL and CLAS
collaborations for helpful discussions and for kindly sharing their data
before publication. W.-T.~C. is grateful to the Universit\"at Mainz for the
hospitality extended to him during his visits. This work was supported in
parts by the National Science Council of ROC under grant
No.~NSC90-2112-M002-032, the Deutsche Forschungsgemeinschaft (SFB 443), and
a joint project NSC/DFG TAI-113/10/0.
\end{acknowledgments}


\end{document}